\documentclass[prl, twocolumn, showpacs]{revtex4}
\usepackage{graphicx}

\begin{document}

\title{\bf Toward a New Phenomenon: Super-\v Cerenkov Radiation}

\author{D. B. Ion}
\affiliation{TH Division, CERN, CH-1211 Geneva 23, Switzerland, and
NIPNE-HH, Bucharest, P.O. Box MG-6, Romania}
\author{M.L. Ion}
\affiliation{Department of Atomic and Nuclear Physics, University of
Bucharest, Romania}

\begin{abstract}
In this letter a new coherent gamma emission mechanism, called
Super-\v Cerenkov radiation, is introduced. The S\v CR is expected
to take place when the charged particle is moving in a medium
with a phase velocity $v_{xph}$ satisfying the super-coherent condition:
$\cos\theta_{SC}=v_{xph}v_{\gamma ph}\leq 1$. The results on an experimental
test of S\v CR in RICH detector are presented.
\end{abstract}


\maketitle

The electromagnetic \v Cerenkov radiation was first observed in
the early 1900's by
the experiments developed by Marie and Pierre Curie when studying
radioactivity
emission. The first deliberate attempt to understand the
phenomenon was
made by Mallet \cite{mal} in 1926. He observed  that the light emitted
by a variety of
transparent bodies placed close to a radioactive source always had
the same bluish-white
quality, and that the spectrum was continuous, not possessing the
line or band structure characteristic of fluorescence. Only the exhaustive
experimental  work, carried out between years 1934-1937 by
P. A. \v Cerenkov \cite{cer},  characterized completely this kind of
radiation. These  experimental data preceded and are fully consistent
with the  classical electromagnetic theory developed by Frank and
Tamm \cite{tam}. So, they showed that charges travelling faster than
the speed of light in a substance with a frequency-independent refractive
index $n_{\gamma}\left(\omega_{\gamma}\right)$ emit coherent radiation  satisfying
the \v Cerenkov coherence  relation (we adopted the system of units $\hbar=c=1$):
\begin{equation}  \label{eq1}
\cos\theta _{C}=v_{\gamma ph}/v_{x}\leq 1
\end{equation}
where $\theta_{C}$ is the angle between the direction of motion and that of
the  electromagnetic wavefront, $v_{x}$ is the speed of the particle in
medium.  A quantum approach of the \v Cerenkov effect by Ginsburg \cite{gins}
resulted  in only minor modification to the classical theory. Now, the
\v Cerenkov radiation (\v CR) is the subject of many studies related to
extension to the nuclear media \cite{db1} as well as to other coherent
particle  emission via \v Cerenkov-like mechanisms \cite{db1}-\cite{db5}. The
generalized  \v Cerenkov-like effects based on four fundamental
interactions  has been investigated and classified recently in \cite{db4}.
In particular, this classification includes the nuclear (mesonic, $\gamma$,
weak boson)-\v Cerenkov-like radiations as well as the high energy
component  of the coherent particle emission via (baryonic, leptonic,
fermionic) \v Cerenkov-like effects.
Recent results on subluminal \v Cerenkov radiation \cite{afan}  as well the
result  on anomalous \v Cerenkov rings \cite{vod} stimulated new theoretical
investigations  \cite{db5} leading us to the discovery  that \v Cerenkov
radiation   is in fact  only a component (low energy component) of a
more general  phenomenon called here Super-\v Cerenkov  radiation (S\v CR)
characterized by the Super-\v Cerenkov  coherence
condition
\begin{equation}  \label{eq2}
\cos\theta _{SC}=v_{x ph} \cdot v_{\gamma ph}\leq 1
\end{equation}
where $v_{x ph}$  and $v_{\gamma ph}\left(\omega\right)$ are
phase velocities of the charged particle and photon, respectively.

{\bf Super-\v Cerenkov radiation.}
While the complete mathematical theory (classical and quantum) can be constructed step
by step as in the case of traditional  \v Cerenkov radiation it is nevertheless
appropriate at this beginning point to explain the basic principles of the Super-\v
Cerenkov effect in qualitative manner. This should enable the reader
to appreciate more fully the interpretation of earlier experiment as well
as the synthesis which includes in a more general and exact form the
recent results on the anomalous and subthreshold \v Cerenkov radiations.
First, we consider that the propagation properties of particles in any medium
(dielectric, nuclear, hadronic, etc.) are changed in agreement with their elastic
scattering with the constituents of that medium. To be more precise, the phase velocity
$v_{xph}\left(E_{x}\right)$ of any particle $x$ (with the total energy
$E_{x}$ and rest mass $M_{x}$ in medium) is modified according to the relation
(we underline again that we work in the units system ${\hbar}=c=1$):
\begin{equation}  \label{eq3}
v_{x ph}\left(E_{x}\right)=\frac{E_{x}}{{\rm Re}\, n_{x}\:\sqrt{E_{x}^{2}-M_{x}^{2}}}
\end{equation}
while the refractive index $n_{x}\left(E_{x}\right)$ in a medium composed from
the constituents ``c'' can be calculated in standard way by using the Foldy-Lax formula
\cite{fol}
\begin{equation}
\label{eq4}
n^{2}_{x}(E_{x})=1+\frac{4\pi \rho }{E_{x}^{2}-M^{2}_{x}}\cdot C(E_{x})
\bar{f}_{xc\rightarrow xc}(E_{x},0^{0})
\end{equation}
where $\rho$ is the density of the constituents ``c'',  $C(E_{x})$ is a coherence
factor  ($C(E_{x})=1$ when the medium constituents are randomly distributed),
$\bar{f}_{xc\rightarrow xc}(E_{x},0^{0})$ is the averaged
forward xc-scattering amplitude.
In order to obtain a simple proof of the Super-\v Cerenkov
condition (\ref{eq2}) we start with the
kinematics of a general (in medium) decay $B_{1}\rightarrow B_{2}+\gamma$
where a photon $\gamma$ [with energy $\omega$, momentum
$k=\omega \cdot {\rm Re}\; n_\gamma(\omega)$,
and refractive index $n_\gamma$] is emitted in a
medium by incident particle $B_1$ [with energy $E_1$, momentum
$p_1={\rm Re}\; n_1(E_1)\:\sqrt{E_1^{2}-M^{2}}$, rest mass
$M$, refractive index $n_1(E_1)$], that itself goes over into a  final
particle [with energy $E_2$, momentum
$p_2={\rm Re}\; n_2(E_2)\:\sqrt{E_2^{2}-M^{2}}$, rest mass
$M$, refractive index $n_2(E_2)$].

The super-\v Cerenkov relation (see Fig. 1a,b):
\begin{equation}  \label{eq5}
\cos\theta _{B_2\gamma}=v_{B_2 ph}\cdot v_{\gamma ph}\leq 1
\end{equation}
can be  easy proved by using  the energy-momentum conservation law  for
the ``decay''  $B_{1}\rightarrow B_{2}+\gamma$  to  obtain
\begin{eqnarray}  \label{eq6}
\cos\theta _{B_2\gamma}=\frac{E_2}{p_2}\cdot \frac{\omega}{k}+
\frac{D_1-D_2-D_\gamma}{2p_2 k} \nonumber \\
\approx v_{B_2 ph}\cdot v_{\gamma ph}
\end{eqnarray}
Here $D_x$, $x\equiv B_1,B_2,\gamma$, are given by $D_x\equiv E_x^2-p_x^2$
in medium. It is important to note that from the
dual super-coherence conditions (\ref{eq5}) two important
generalized \v Cerenkov-like limits follow: the
photon \v Cerenkov limit: $v_{\gamma ph}\leq v_{B_1ph}^{-1}$ (see Fig.1a), and Coherent
B-\v Cerenkov-like limit: $v_{B_2 ph}\leq v_{B_1ph}^{-1}$ (see Fig. 1b).

\begin{center}
\noindent
\begin{minipage}{8.5cm}
\includegraphics[width=8cm]{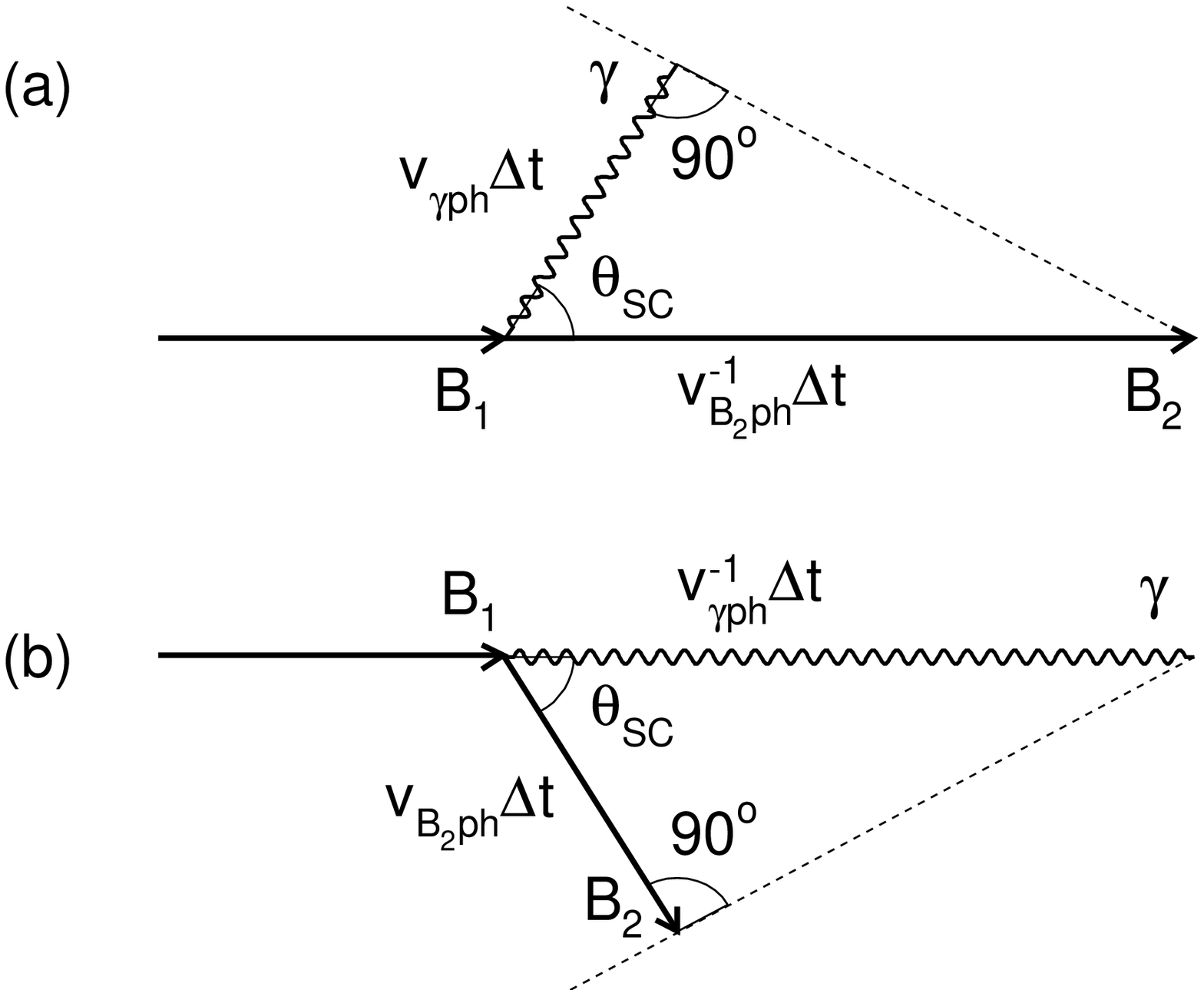}\\

{{\bf Figure 1:} \footnotesize
Schematic description of  Super-\v Cerenkov effect: (a) Cerenkov
radiation sector: $\cos\theta _{SC}=v_{\gamma ph}\cdot v_{B_1 ph}\leq 1$;
(b) \v Cerenkov-like bremsstrahlung radiation sector:
$\cos\theta _{SC}=v_{B_1 ph}\cdot v_{B_2 ph}\leq 1$. }
\end{minipage}
\end{center}

To be more specific
let us consider a charged particle (e.g. $e^\pm$, $\mu^\pm$, $p$,  etc.) moving
in a (dielectric, nuclear or hadronic) medium and to explore the
$\gamma-$coherent emission via Super-\v Cerenkov mechanism in that media.
In the case of dielectric medium the \v Cerenkov radiation is a well established
phenomenon widely used in physics and technology. Also, experimentally, the
high energy $\gamma-$emission via coherent bremsstrahlung is well known as a
channelling effect in many crystals. So, an important problem now is if  the high
energy component of the Super-\v Cerenkov phenomenon can be identified with the
coherent bremsstrahlung radiation. Hence, more experimental and theoretical
investigations are needed since the usual \v Cerenkov radiation and Coherent
bremsstrahlung radiations can be described in a unified way via the Super-\v
Cerenkov as two-component generalized \v Cerenkov-like effects.

Now, the subthreshold rings observed experimentally can also be interpreted
as Super-\v Cerenkov signatures since in the \v CR-sector $\theta_{SC}
\equiv\theta_{2\gamma}=\theta_{1\gamma}$(Fig.1a) the number of photons emitted
in the intervals $(x, x+dx)$, $(\omega,\omega+d\omega)$ in an
nonabsorbent medium will be given by
\begin{equation} \label{eq7}
\frac{d^2N}{dxd\omega }=\alpha Z_{B_1}^2 \sin^2\theta_{1\gamma}
=\alpha Z_{B_1}^2 \left( 1-v^{2}_{\gamma ph}v^{2}_{B_{1}ph}\right)
\end{equation}
where $\alpha=1/137$ is the fine structure constant and $Z_{B_1}$ the
electric charge of the $B_1$ particle. Indeed,  it is easy to see
that the Super-\v Cerenkov coherence condition (\ref{eq5}) includes in a
general and exact form the subthreshold \v Cerenkov-like
radiation \cite{afan}  since: $v_{B_1}^{thr}(SCR)=v_{B_1}^{thr}(CR)/
{\rm Re}n_{B_1}$ for  ${\rm Re}n_{B_1}\geq 1$.
Moreover, the anomalous \v Cerenkov rings observed recently \cite{vod}
at SPS accelerator at CERN Pb-beam can be considered as experimental
signature of the high-energy  component of the Super-\v Cerenkov effect
(see Fig. 1b) since both $\gamma$ and $Pb$ after emission of high energy
photon can produce secondary (anomalous) rings. The principal signatures
of the Super-\v Cerenkov radiation  for these two limiting
sectors  are given in Table 1.

\begin{center}
\noindent
{\bf Table 1:} The  S\v CR  main predictions.
\nopagebreak

{\footnotesize
\noindent \begin{tabular}{|p{2mm}|p{1.3cm}|p{2.9cm}|p{2.7cm}|} \cline{1-4}
  & Name & \parbox[c]{2.4cm}{\centering (S\v CR) Low $\gamma$-energy sector (Fig.1a)} &
 \parbox[c]{2.4cm}{\centering (S\v CR) High $\gamma$-energy sector (Fig.1b)} \\ \cline{1-4}
1 &  \parbox[c]{1.4cm}{coherence relation} & $v_{\gamma ph}\cdot v_{xph}\leq 1$ &
$v_{\gamma ph}\cdot v_{xph}\leq 1$ \\ \cline{1-4}
2 &  \parbox[c]{1.4cm}{coherence angle} &  \parbox[c]{2.7cm}{\centering
$\cos\theta _{SC}=v_{\gamma ph}\cdot v_{1ph}$
since $\theta_{SC}\equiv \theta_{2\gamma}\approx \theta_{1\gamma}$} &
 \parbox[c]{2.7cm}{\centering $\cos\theta _{SC}=v_{1 ph}\cdot v_{2ph}$
since $\theta_{SC}\equiv \theta_{2\gamma}\approx \theta_{12}$} \\ \cline{1-4}
3 &  \parbox[c]{1.4cm}{threshold velocity} &  \parbox[c]{2.7cm}{\centering
$v_{xthr}(SC)=\frac{1}{n_1n_\gamma}$} &
 \parbox[c]{2.7cm}{\centering $v_{xthr}(SC)=\frac{1}{n_1n_2}$} \\ \cline{1-4}
4 &  \parbox[c]{1.4cm}{maximum emission angle} &  \parbox[c]{2.7cm}{\centering
$\theta_{SC}^{max}=\arccos\frac{1}{n_1n_\gamma}$} &
 \parbox[c]{2.7cm}{\centering $\theta_{SC}^{max}=\arccos\frac{1}{n_1n_2}$} \\ \cline{1-4}
5 &  \parbox[c]{1.4cm}{\vspace{1mm} spectrum \\} &  \parbox[c]{2.9cm}{\centering
$\frac{dN_{\gamma}}{d\omega }=\alpha LZ^2 \sin^2\theta_{SC}$} &
 \parbox[c]{2.7cm}{\centering $\frac{dN_{\gamma}}{d\omega }=LZ^2/2\;^*)$} \\ \cline{1-4}
6 &  \parbox[c]{1.4cm}{polari-\\zation} & 100\% ($\vec{e}||Q$) &
100\% ($\vec{e}\bot Q$)  $^{**})$\\ \cline{1-4}
\end{tabular}

\noindent
$^*$) L is the particle  path length in the medium, \\
$^{**}$) Q is the  S\v CR decay plane.
}
\end{center}

All these results can be proved in a  quantum theory of the Super-\v Cerenkov
effect. Here, we give only some final results. So, just as in the quantum
theory of \v CR, the same interaction Hamiltonian $H_{fi}$ with some modifications
of the source fields in medium can also describe the coherent $\gamma-$emission
in all sectors. Then, it is easy to see that intensity of  the
Super-\v Cerenkov radiation can be given in the form
\begin{equation} \label{eq8}
\frac{d^2N}{dtd\omega }=\frac{\alpha Z^2}{v_1}
\frac{1}{|n_{B_1}|^2|n_{B_2}|^2|n_{\gamma}|^2}
\frac{k}{\omega}\frac{dk}{d\omega}\,S\cdot\Theta\left(1-\cos \theta_{SC}\right)
\end{equation}
where the spin factor $S$, for a two-body spin ($1/2^+\rightarrow \gamma+
1/2^+$) electromagnetic ``decay'' in medium, is given by
\begin{eqnarray}  \label{eq9}
S\equiv \frac{(E_1+M)(E_2+M)}{4E_1E_2}\cdot
\biggl[ \frac{p_1^2}{E_1+M}+\frac{p_2^2}{E_2+M}+ \nonumber\\
+2\frac{(\vec{e}_k.\vec{p}_1)(\vec{e}_k.\vec{p}_2)-
(\vec{e}_k \times \vec{p}_1)(\vec{e}_k \times \vec{p}_2)}{(E_1+M)(E_2+M)}\biggr]
\end{eqnarray}

Now,  one can see that $\Theta\left(1-\cos \theta_{SC}\right)$ Heaviside function
is 1 in two (or many)
physical regions defined by the constraint: $\cos \theta_{2\gamma}
\approx v_{\gamma ph}(\omega)v_{2ph}(E_2)\leq 1$. So, the low $\gamma-$energy
sector is that where $\theta_{SC}\equiv \theta_{2\gamma}\approx \theta_{1\gamma}$,
while the  high $\gamma-$energy sector is that
where $\theta_{SC}\equiv \theta_{2\gamma}\approx \theta_{12}$.
Hence, the low $\gamma-$energy sector
can be identified as extended \v Cerenkov region
[$v_{1ph}^{-1}(E_1)\geq v_{\gamma ph}(\omega)$] which  includes
the medium modifications on
the propagation properties of charged particle, while high $\gamma-$energy
sector can be identified as extended \v Cerenkov-like mechanism to the
charged particle in the sense that the source spontaneously
decays in a high-energy photon according to a \v Cerenkov-like
relation: $v_{1ph}^{-1}(E_1)\geq v_{2ph}(E_2)$.
The spin factor $S$ in Eqs. (\ref{eq8})-(\ref{eq9}) is defined just as
in the usual quantum theory  of CR but with
the particle's  momenta  $P_i$, $i=1,2$  considered in medium.
The vector $\vec{e}_k$ is the photon polarization  for a given photon momentum
$\vec{k}$.  Now, if for a given $\vec{k}$ one chooses two orthogonal
photon spin polarization directions, corresponding to a polarization
vector perpendicular and parallel to the plane given by $\vec{p}_1$ and $\vec{k}$,
respectively, the corresponding spin factors are given by
\begin{eqnarray}  \label{eq10}
S^\bot = \frac{(E_1+M)(E_2+M)}{4E_1E_2}\cdot
\biggl( \frac{\vec{p}_1}{E_1+M}-\frac{\vec{p}_2}{E_2+M}\biggr)^2
\end{eqnarray}
and
\begin{eqnarray}  \label{eq11}
S^{||} =v_1 {\rm Re}\, n_1\; v_2q{\rm Re}\, n_2\;
\sin\theta_{1\gamma}\;\sin\theta_{2\gamma}
\end{eqnarray}

Therefore, the relations (\ref{eq8})-(\ref{eq11}) includes in a general and unified way
all the main predictions of the Super-Cerenkov radiation from which the
results from Table 1 are obtained as two particular limiting cases.
In the SC-Low energy sector we have 100\% linear polarization $\vec{e}||Q$,
while in the SC-High energy sector we get 100\% linear polarization $\vec{e}\bot Q$,
where $Q$ is the ``S\v CR-decay'' plane.

{\bf Experimental tests of  Super-\v Cerenkov effect.}
\v Cerenkov radiation is extensively used in experiments for counting
and identifying relativistic particles in the fields of elementary
particles, nuclear physics and astrophysics. A spherical mirror focuses
all photons emitted at \v Cerenkov angle along the particle trajectory
at the same radius on the focal plane. Photon sensitive detectors placed
at the focal plane detect the resulting ring images in a  Ring Imaging
\v Cerenkov (RICH) detector. So, RICH-counters are used for identifying
and tracking  charged particles. \v Cerenkov rings formed on a focal
surface of the RICH provide information about the velocity and the
direction of a charged particle passing the radiator. The particle's
velocity is related to the Cerenkov angle $\theta_C$ (or to the
Super- \v Cerenkov  $\theta_{SC}$ ] by the relation (1) (or (2), respectively).
Hence, these angles are determined by measuring the radii of the rings
detected with the RICH.  In ref.  \cite{deb}  a $C_4F_{10}Ar(75:25)$
filled RICH-counter read out  was used
for measurement of the \v Cerenkov ring radii. Fig. 2a  shows the
experimental values of the ring radii of  electrons, muons, pions and
kaons measured  in the active area of this RICH-detector.
The saturated light produced from
electrons was a decisive fact  to take  an
index of refraction $n_\gamma=1.00113$ for the radiator material.
The absolute
values for excitation curves of electron, muon, pion and kaon, shown
by dashed curves in Fig. 2a,  was obtained by using this value of refractive
index in formula: $r_C(p)=(R/2)\tan \theta_C(p)$. The solid curves show  the
individual best fit of the experimental
ring radii with eqn. $r_{SC}(p)=(R/2)\tan \theta_{SC}(p)$(see Table 1).
For the particle refractive index we used the parametrization
\begin{eqnarray}  \label{eq15}
n_x^2(p)=1+a^2/p^2,\;\;v_x=p/\sqrt{p^2+m^2}
\end{eqnarray}

\begin{center}
\noindent {\bf Table 2:}  The best fit parameters of experimental ring radii with the
Super-\v Cerenkov prediction.
\noindent
\begin{tabular}{|c|r|r| r|}
\hline
Particle & Mass (MeV)& $10^3\cdot a^2$ (GeV/c)&$\chi^2/n_{dof}$\\ \hline
$e$   &   0.511 & -0.081$\pm$0.101 & 0.468 \\
$\mu$ & 105.658 &  1.449$\pm$0.098 & 3.039 \\
$\pi$ & 139.570 &  2.593$\pm$0.167 & 0.234 \\
$K$   & 493.677 & 21.140$\pm$2.604 & $<10^{-14}$ \\
\hline
\end{tabular}
\end{center}

We fitted all the 18 experimental data on the ring radii from ref.\cite{deb}
with our Super-\v Cerenkov prediction formula
\begin{eqnarray}  \label{eq14}
r_{SC}(p/m)=\frac{R}{2}\tan \theta_{SC}=\frac{R}{2}
(n_\gamma^2n_x^2v_x^2-1)^{1/2}=  \nonumber\\
=\frac{R}{2}\biggl[n_\gamma^2\frac{(p/m)^2+(a/m)^2}{(p/m)^2+1}-1\biggr]^{1/2}
\end{eqnarray}
and we obtained the following  consistent  result (see Fig. 2b).
The best fit   parameters are as follow: $(a/m)^2=0.12109\pm 0.00528$
and  $\chi^2/n_{dof}=1.47$, where $n_{dof}=16$
is the number of degree of freedom (dof). The $r_{SC}(p/m)$ scaling function
(\ref{eq14}) together with  all experimental data on the ring radii of the electron,
muon, pion and kaon,  are plotted as a function of the scaling variable
(p/m) in Fig. 2b.

{\bf Conclusions.} In this letter a description  of a new dual coherent
particle production
mechanism, called Super-\v Cerenkov mechanism  (S\v CR), is presented:

{\bf(i)} The S\v CR-phenomenon,  as generalized two-component \v Cerenkov-like effect,  can
be viewed as a continuous two body decays $B_1\rightarrow \gamma+B_2$
in medium and is expected to take
place when the phase velocities of the emitted photon $v_{\gamma ph}$
and that of particle
source $v_{B_1ph}$ satisfy the dual super-coherence condition:
$v_{\gamma ph}\cdot v_{B_1ph}\leq 1$. It is shown that the S\v CR
includes in a general and exact form two coherent limiting phenomena: coherent
\v Cerenkov emission (see Fig.1a) and a  \v Cerenkov-like effect for the charged
particles (see Fig. 1b, Table 1).

{\bf(ii)} The results on  experimental test of the super-coherence conditions are
presented in Fig. 2a,b. These S\v CR-predictions are verified experimentally
with high accuracy $\chi^2/n_{dof}=1.47$ (see Fig.2b)
by the data \cite{deb} on the \v Cerenkov ring
radii of electron, muon, pion and  kaon, all measured with RICH detector.

\begin{center}
\begin{minipage}{8.5cm}
\includegraphics[width=8cm]{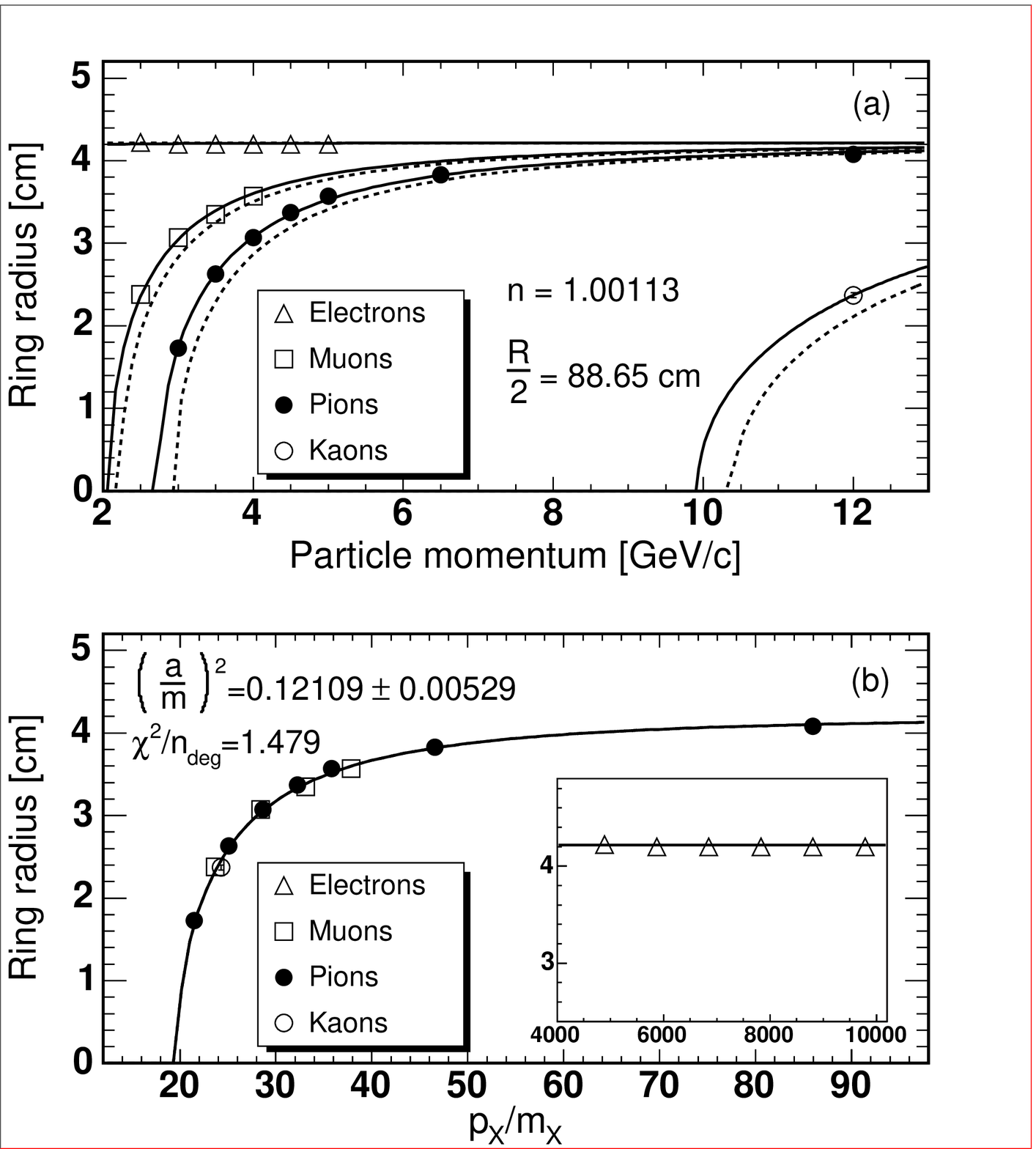}\\
{{\bf Figure 2:} \footnotesize
\v Cerenkov ring radii of the particles $e$, $\mu$, $\pi$, $K$
obtained in Ref. \cite{deb}  with    $C_4F_{10}Ar-$RICH detector are
compared with: (a) the  theoretical Super-\v Cerenkov prediction
(solid curves) as well as with \v Cerenkov  values (dashed curves).
(b) The scaling Super-\v Cerenkov radius $r_{SC}$ (\ref{eq14}) as a
function of (p/m) (see text).}
\end{minipage}
\end{center}

{\bf(iii)} We shown that the Super-\v Cerenkov phenomenon can explain not only
subthreshold CR \cite{afan} but also the observed secondary rings (or anomalous
\v Cerenkov radiation) at CERN SPS accelerator \cite{vod}.

{\bf(iv)} The influence of medium on the particle propagation properties  is
investigated  and the refractive properties of electrons, muons, pions,
in the radiator  $C_4F_{10}Ar$ are obtained.
The refractive indices for this radiator at $p_{lab}=1GeV$ are as follows:
$n_{\mu}=1.001449\pm 0.000098$, $n_{\pi}=1.0012593\pm 0.000167$,
$n_{K}=1.0214\pm 0.0026$, $n_{p}=1.1066\pm 0.046$. So, we proved that the refractive indices  of
the particles in medium are also very important for the RICH detectors, especially at
low and intermediate energies.

{\bf Acknowledgments.} We would like to thank G. Altarelli for fruitful
discussions. One of the authors (D.B.I.) would like to thank TH Division
for hospitality during his stay at CERN. This work was supported by CERES Projects
C2-86-2002, C3-13-2003.

\end{document}